\documentclass[twocolumn,amsmath,amssymb,fleqn,prb]{revtex4-2}

\usepackage{cancel}
\usepackage{graphicx}%
\usepackage{bm}
\usepackage{subfigure}

\begin{document}

\title{On the relation between thermodynamical and statistical entropy: \\ The origin of the $N!$}
\author{S. Di Matteo$^1$}
\affiliation{$^1$ Univ Rennes, CNRS, IPR (Institut de Physique de Rennes) - UMR 6251, F-35708 Rennes, France}

\date{\today}

\begin{abstract} 
The division by $N!$ in the expression of statistical entropy is usually justified to students by the statement that classical particles should be counted as indistinguishable. Sometimes, {\it quantum} indistinguishability is invoked to explain it. 
In this paper, we try to clarify the issue starting from Clausius thermodynamical entropy and deriving from it Boltzmann statistical entropy for the ideal gas. This approach appears interesting for two reasons: Firstly, it provides a direct heuristic link between thermodynamical and statistical expressions of entropy that is missing in the usual approach. In second place, it explicitly reminds that also statistical entropy is defined with respect to a reference state, chosen to be the quantum-mechanical ($T=0$) ground state of the $N$-particles in a box. Both factors $h^{3N}$ and $N!$ at the denominator of the statistical entropy are a consequence of the quantum nature of the reference state: in particular, the $N!$ is not related to the particle identity, but to the identity of the boundary conditions on the quantum-mechanical ground-state momenta, and to a general statistical maximization principle.  
The introduction of a reference state also allows to reinterpret the standard case of two mixing gases. 

\end{abstract}

\maketitle

\section{Introduction}

A typical introduction\cite{huang,chandler} to classical statistical physics of an isolated system in undergraduate courses is based on the three postulates: {\it (1)} Ergodicity (to equate time-averages to configuration-averages in phase space); {\it (2)} Equiprobability of the accessible phase-space volume; {\it (3)} Definition of statistical entropy for $N$ particles of internal energy $U$ in a volume $V$ through the revised\cite{noteB} Boltzmann's expression\cite{sw1,comment}: 

\vspace{-0.1cm}
\begin{equation}
\hspace{18mm}S(U,V,N)=k_B\ln\frac{\Omega(U,V,N)}{h^{3N}N!}
\label{entro}
\end{equation}
\vspace{-0.1cm}

\noindent where $k_B$ is Boltzmann's constant, $h$ Planck's constant and $\Omega(U,V,N)=\int_{{\cal A}(p,q)} (dp)^{3N}(dq)^{3N}$ the available phase-space volume, with ${\cal A}(p,q)$ the phase-space region defined by $U\leq {\cal H}(q,p)\leq U+\Delta$. Here ${\cal H}(q,p)$ is the Hamiltonian of the $3N$ position coordinates $q_i$ and the $3N$ quantities of movement $p_i$ and $\Delta$ a constant energy defining the width of the phase-space shell. 

However, Eq. (\ref{entro}) appears to students a bit from nowhere: why should a phase-space volume be related to the quantity $\frac{\delta Q}{T}_{|{\rm rev}}$? why does the Planck's constant appear in a classical expression? and why should we divide by $N!$ - a point that, though already clear to the 'pre-quantistic' Boltzmann\cite{sw1} and Gibbs,\cite{gibbs} is often\cite{huang,kubo} associated to the {\it quantum} indistinguishability of the particles?

The aim of this article is to answer the above three questions by showing that, for an ideal gas, Eq. (\ref{entro}) can be derived heuristically from the Clausius' expression of the thermodynamical entropy, $dS=\frac{\delta Q}{T}_{|{\rm rev}}$, if we recognize that, as it is the case for thermodynamical entropy, also statistical entropy should be defined with respect to a reference state. As we shall see, such a reference state is the quantum-mechanical ground state of the system, to which we associate the reference value $S(T=0)=0$ by the third principle of thermodynamics and which leads to both N! and $h^{3N}$. 

The analysis of the role of the $N!$ in Eq. (\ref{entro}) was also stimulated by some discussions in the recent literature\cite{sw1,dieks,sw2,sw3}.
In particular, R.H. Swendsen proposed a solution\cite{sw1} to include the description of colloids by imposing entropy extensivity\cite{paulicomm}. His method is logically irreproachable, though it seems in contradiction with the classical limit of quantum statistical mechanics, where the $N!$ term appears to be a consequence of the quantum indistinguishability of the particles (see, e.g., Ref. [\onlinecite{kubo}] for a detailed derivation). As a student, I was indeed taught that the division by $N!$ in Eq. (\ref{entro}) is a consequence\cite{huang,kubo} of {\it quantum} indistinguishability - yet, this is not correct, as shown below. 
For the sake of clarity, we remind that $N$ particles are said identical if the Hamiltonian is invariant with respect to the permutation of their $N$ coordinates. The corresponding quantum states (either bosonic or fermionic) are said indistinguishable: the wave function is invariant or changes sign after such a permutation\cite{bransden}. In this work, dealing only with ideal gas of classical point particles, particles are identical if they have the same mass.

In the light of some literature\cite{dieks}, the first point to clarify is why Eq. (\ref{entro}), with $N!$ at the denominator, is correct. In order to avoid too philosophical digressions, the best pedagogical choice is to stick to experimental results, presented in Section II, to validate the Sackur-Tetrode's formula,\cite{sackur,sackurbis,tetrode,grimus,exposito} equivalent to Eq. (\ref{entro}) in the thermodynamic limit of the ideal gas. This experimental validation allows to reject other extensive expressions of the entropy and clarifies why {\it quantum} indistinguishability is {\it not} at the origin of the $N!$. It serves also as a guide to our main result in Section III, i.e., the derivation of a reference state for statistical entropy from the analogy with thermodynamical entropy: recognizing that, necessarily, $N!$ and $h^{3N}$ are features of the reference state allows analyzing them from a different perspective.  
In Section IV, we shall use our scheme to reinterpret the so-called entropy of mixing, which is not determined by mixing itself, but by volume variations, as already suggested long ago\cite{bennaim}. In Section V we draw our conclusions, lingering on the meaning of {\it quantum} vs. {\it statistical} indistinguishability.

\section{Sackur-Tetrode equation}

In 1912, Sackur\cite{sackur,sackurbis} and Tetrode\cite{tetrode}, starting from Eq. (\ref{entro}), independently found the following expression for the entropy of an ideal gas (a detailed derivation is in [\onlinecite{grimus}]):
\vspace{-0.1cm}
\begin{align}
S_{\rm ST}(U,V,N) = k_B N \left\{\ln\left[\left( \frac{4\pi mU}{3h^{2}N} \right)^{\frac{3}{2}}\frac{V}{N}\right] +\frac{5}{2} \right\}
\label{sackur1}
\end{align}
\vspace{-0.1cm}

Actually, Sackur, in his first paper, did not count the particles correctly\cite{sackur,grimus} with the $N!$, but considered the volume per particle, $\frac{V}{N}$. His first expression for the entropy was\cite{refsac}:  

\vspace{-0.1cm}
\begin{align}
S_{w1}(U,V,N) = k_B N \left\{\ln\left[\left( \frac{4\pi mU}{3h^{2}N} \right)^{\frac{3}{2}}\frac{V}{N}\right] +\frac{3}{2} \right\}
\label{sackur2}
\end{align}
\vspace{-0.1cm}

\noindent with the term $\frac{3}{2} k_BN$ instead of $\frac{5}{2} k_BN$ of Eq. (\ref{sackur1}). For systems with fixed number of particles the two expressions only differ by a constant value, $Nk_B$, and both describe an extensive entropy. So we might accept both equations as valid\cite{noteMa}, with the idea that only entropy differences are meaningful. 
Yet, even for fixed values of $N$, experimental data show us that only Eq. (\ref{sackur1}) is valid if we accept the third principle of thermodynamics. 
To understand why, consider how experimental entropy, $S^e(T)$, is obtained for a gas\cite{booksthermo}. First, the heat capacity at constant pressure, $C_p(T)$, is measured by calorimetric methods at several temperatures. Then, the equation $S^e(T)-S^e(T_0)=\int_{T_0}^T d\tau \frac{C_p(\tau)}{\tau}$ allows to {\it calculate} entropy variations by numerical integration. If a phase transition is passed through, the latent heats (of fusion, $|\Delta S^e_{\rm fus}|$, and vaporization, $|\Delta S^e_{\rm vap}|$) should be added. 
Consider the actual values for natural krypton: by imposing the third principle of thermodynamics $S^e(T=0)=0$, the experimental entropy for Kr in the gas state at a temperature immediately above $T_{\rm vap}^{\rm Kr}$ is then obtained as: 

\vspace{-0.1cm}
\begin{align}
S^e(T) &=\!\! \int_0^{T_{\rm fus}^{\rm Kr}}\!\!\!\!\! d\tau \frac{C_p(\tau)}{\tau} +|\Delta S^e_{\rm fus}| +\!\!\int_{T_{\rm fus}^{\rm Kr}}^{T_{\rm vap}^{\rm Kr}}\!\!\! d\tau \frac{C_p(\tau)}{\tau}+|\Delta S^e_{\rm vap}| \nonumber \\
&= \Delta S^e_{\rm sol}+|\Delta S^e_{\rm fus}|+\Delta S^e_{\rm liq}+|\Delta S^e_{\rm vap}|
\label{expent}
\end{align}
\vspace{-0.1cm}

As we cannot experimentally reach $T=0$ K, the lowest temperature region is approximated theoretically by Debye's law for solids. In the case of Kr, by taking Debye's temperature\cite{DebyeT} $\theta_D=71.9$ K, we obtain the entropy per mole\cite{extens} $s^e(1\,{\rm K})\equiv s_{\rm Debye}(1\,{\rm K})=0.002$ Jmol$^{-1}$K$^{-1}$, with $s=\frac{S}{n}$. We remark that this numerical value is absolutely negligible, as below the experimental uncertainty for $s_e(T)$.
From this temperature on, it is possible to integrate numerically the experimental data\cite{exp12} on the heat capacity at constant pressure. We use the results of Ref. [\onlinecite{exposito}]: for the solid region (from 1 K to $115.77$ K) $\Delta s^e_{\rm sol} = 53.23$ Jmol$^{-1}$K$^{-1}$ and in the liquid phase (from 115.77 K to 119.81 K) $\Delta s^e_{\rm liq} = 1.50$ Jmol$^{-1}$K$^{-1}$. Finally, from the Handbook of Chemistry and Physics\cite{handbook} at the melting point (from the latent heat of fusion) $|\Delta s^e_{\rm fus}| = 14.215$ Jmol$^{-1}$K$^{-1}$ and at the boiling point (from the latent heat of vaporization) $|\Delta s^e_{\rm vap}| = 75.834$ Jmol$^{-1}$K$^{-1}$. 
The sum of all these terms leads to the entropy at the vaporization point: $s^e(T_{\rm vap}^{\rm Kr})=144.8\pm 0.2$ Jmol$^{-1}$K$^{-1}$, where the uncertainty is considered on the last digit of the latent heats\cite{handbook} and is given with one digit. 

If we now evaluate the theoretical expression\cite{virial} of the ideal gas $s^{\rm th}(T)$ through either Eq. (\ref{sackur1}) or  Eq. (\ref{sackur2}), we get: 

1) by using Eq. (\ref{sackur1}), $s_{\rm ST}(T_{\rm vap}^{\rm Kr})=145.0$ Jmol$^{-1}$K$^{-1}$, in very good agreement\cite{notexpo} with $144.8 \pm 0.2$ Jmol$^{-1}$K$^{-1}$.

2) by using Eq. (\ref{sackur2}), $S_{w1}(T_{\rm vap}^{\rm Kr})=136.7$ Jmol$^{-1}$K$^{-1}$, in relatively poor agreement with the experimental value.

We remark that Sakur and Tetrode used their equation in the opposite way, to fit the value of the Planck's constant from the experimental data of entropy, and could obtain a good agreement only by using Eq. (\ref{sackur1}). Indeed, using Eq. (\ref{sackur2}), Sackur initially found a value of the Planck's constant wrong by the factor $\sqrt[3]{e}$. 
We can therefore conclude that the division by $N!$ in Eq. (\ref{entro}) is not important to 'just' require entropy extensivity, as the latter could have been obtained also by replacing $\frac{V^N}{N!}\simeq \left(\frac{Ve}{N}\right)^N$ with $\left(\frac{V}{N}\right)^N$, the volume per particle\cite{sackur,noteMa}. 
It is important also because it allows measuring the Planck's constant, by selecting Eq. (\ref{sackur1}) among all possible extensive entropies. 
We shall be back on this point in the Conclusions. Now, compare the previous case of Kr with the analogous excellent agreement of Eq. (\ref{sackur1}) with the experimental\cite{exposito} values for Ar: $129.2$ Jmol$^{-1}$K$^{-1}$ vs. $128.8\pm 0.3$ Jmol$^{-1}$K$^{-1}$. 
Ar is $99.6\%$ mono-isotopic: it can be considered as made of identical particles. On the other side, Kr gas is a mixing of at least 5 isotopes: it is not made of identical particles (if we limit to relative abundance higher than $1\%$, we have $^{84}$Kr ($\gamma_1\simeq 56.99\%$), $^{86}$Kr ($\gamma_2\simeq 17.28\%$), $^{82}$Kr ($\gamma_3\simeq 11.59\%$), $^{83}$Kr ($\gamma_4\simeq 11.50\%$) and $^{80}$Kr ($\gamma_5\simeq 2.29\%$)). Had we calculated $s_{\rm ST}(T_{\rm vap}^{\rm Kr})$ by replacing $N!$ by $\Pi_{i=1}^5 (\gamma_i N)!$ (in order to consider only the permutations belonging to each of the 5 classes of identical particles), we would have obtained the extra entropy contribution $R\sum_{i=1}^5 \gamma_i\ln\frac{1}{\gamma_i}\simeq 1.21R\simeq 10.1$ Jmol$^{-1}$K$^{-1}$, in clear disagreement with the experimental value. In order to agree with the experimental data, we must divide by $N!$ in both Ar and Kr cases, i.e., for identical and non-identical particles, {\it ergo} the $N!$ is not determined by the {\it quantum} indistinguishability of the particles. We shall see theoretically why in Section III.

\section{From thermodynamical to statistical entropy in five acts.}

The aim of this long section is to detail the heuristic derivation of Eq. (\ref{entro}) for the ideal gas starting from Clausius' entropy (Eq. (\ref{thermoentro1}) below). As Clausius' entropy is extensive, the idea is to understand at which point in the derivation, one is forced to introduce $N!$ to recover the extensivity and why it was lost. We remark again that, in the classical approach to statistical mechanics, Eq. (\ref{entro}) is postulated (together with ergodicity and equal apriori probability, in the microcanonical ensemble) and its ability to describe thermodynamical entropy is verified aposteriori. Here, we try to suggest a different pedagogical approach\cite{noteXX}.

{\it Act I - Thermodynamics: the safe harbor.} Consider for simplicity a mono-atomic ideal gas with fixed number of particles, $N$. The first principle of thermodynamics is: $dU = TdS -PdV-\mu dN \overset{dN=0}{=} TdS -PdV$. For an ideal gas, the equation of state and the expression of the internal energy are, respectively: $PV=Nk_BT$ and $U=\frac{3}{2}Nk_BT$. The factor $\frac{3}{2}$, critical for the subsequent analysis, can be obtained theoretically from classical mechanics\cite{ma2} and does not require\cite{remark1} Eq. (\ref{entro}).

The first principle leads to: $dS = \frac{1}{T}dU+\frac{P}{T}dV$. From the equation of state we get $\frac{P}{T}=\frac{Nk_B}{V}$ and the relation of temperature and internal energy gives $\frac{1}{T} = \frac{\frac{3}{2}Nk_B}{U}$. In this way, we obtain the thermodynamical entropy variation: $dS(U,V,N) = \frac{\frac{3}{2}Nk_B}{U}dU+\frac{Nk_B}{V}dV$, as a function of $U$, $V$ and $N$. By integrating once at constant volume and then at constant internal energy from an initial state characterized by $(U_0,V_0,N)$ to a final state characterized by $(U,V,N)$, we get: 

\begin{align}
\vspace{-0.2cm}
\Delta S(U,V,N) & = \frac{3}{2}Nk_B\ln\left(\frac{U}{U_0}\right) + Nk_B\ln\left(\frac{V}{V_0}\right) + S_0(N) \nonumber \\
& = k_B \ln\left[ \left(\frac{U^{\frac{3}{2}}V}{U_0^{\frac{3}{2}}V_0}\right)^N \right]  + S_0(N)
\label{thermoentro1}
\end{align}

\noindent where $S_0(N)$ is the integration 'constant' (with respect to $U$ and $V$). Two remarks on this known result: 

1) The thermodynamical entropy variation, expressed by Eq. (\ref{thermoentro1}), is a state function that depends on the parameters of both the initial $(U_0,V_0,N)$ and final $(U,V,N)$ states of the transformation. Therefore if we consider entropy as a function of $(U,V,N)$, its numerical value is always related to a reference state $(U_0,V_0,N)$ that we need to specify, i.e., $\Delta S(U,U_0,V,V_0,N)$ would be a more correct notation for Eq. (\ref{thermoentro1}). 

2) Integrating the Clausius formula $dS=\frac{\delta Q}{T}$ at constant $N$ does not allow to determine the full dependence on the number of particle: in principle, $S_0$ is not a constant,\cite{paulicomm} but a function of $N$. Imposing extensivity to both initial and final states leads to $S_0(N)=\alpha N$, with $\alpha$ a non-negative constant. As entropy is a state function, $\Delta S(U_0,V_0,N)=0$, and therefore Eq. (\ref{thermoentro1}) implies that $S_0(N)=0$. We remark that our analysis is different from the one presented in Refs. [\onlinecite{pauli,jaynes}], where no explicit dependence on the initial state $(U_0,V_0)$ is given. Here we demand the extensivity in the form $\Delta S(qU,qU_0,qV,qV_0,qN)=q\Delta S(U,U_0,V,V_0,N)$, whereas Refs. [\onlinecite{pauli,jaynes}], not having a reference state, demanded $S(qU,qV,qN)=qS(U,V,N)$: therefore their ${\tilde{S}}_0(N)$ (different from the $S_0(N)$ in Eq. (\ref{thermoentro1})) had to provide the lost extensivity associated to the term\cite{temp} $\ln V_0^{-1}$. 

\vspace{0.3cm}

{\it Act II - Classical Mechanics: sailing along the coast.} Consider now the mechanical description of an ideal gas of $N$ particles of identical mass $m$, whose internal energy is entirely kinetic: $U= \frac{1}{2m}\sum_{j=1}^{3N} p_{j}^2$, with $p_j$ the $j^{\rm th}$-component of the quantity of movement, where the label $j$ runs over all $3N$ degrees of freedom.
We introduce\cite{notex} the variance of the distribution of the $p_{j}$:

\begin{align}
\vspace{-0.2cm}
\hspace{0,5cm} (\Delta p_{\sigma})^2\equiv\frac{1}{3N}\sum_{j=1}^{3N} p_{j}^2-\left(\frac{1}{3N}\sum_{j=1}^{3N} p_{j}\right)^2=\frac{2mU}{3N}
\label{psigma}
\end{align}

\noindent where the last equality follows from the gas being globally at rest (the average value of $p_j$ over all particles is zero). So $U= \frac{3N (\Delta p_{\sigma})^2}{2m}$. 
Analogously $U_0= \frac{3N (\Delta p_{0\sigma})^2}{2m}$. For simplicity of notation, we temporarily suppose that the box is cubic in both states, with volumes $V=(\Delta q)^3$ and $V_0=(\Delta q_0)^3$.

After these manipulations, Clausius' entropy becomes: 

\begin{align}
\vspace{-0.1cm}
S(U,V,N) & = k_B \ln\left[ \left(\frac{\cancel{\left(\frac{3N}{2m}\right)^{\frac{3}{2}}}(\Delta p_{\sigma})^3(\Delta q)^3}{\cancel{\left(\frac{3N}{2m}\right)^{\frac{3}{2}}}(\Delta p_{0\sigma})^3(\Delta q_0)^3}\right)^N \right]  \nonumber \\
& = k_B \ln\left[ \left(\frac{\Delta p_{\sigma}\Delta q}{\Delta p_{0\sigma}\Delta q_0}\right)^{3N} \right] 
\label{thermoentro2}
\end{align}

We remark that Eq. (\ref{thermoentro2}) has the same physical content as the entropy of Eq. (\ref{thermoentro1}) with $S_0(N)=0$, as no approximations have been performed in moving from one to the other. So, Eq. (\ref{thermoentro2}) tells us that thermodynamic entropy can be written (in units $k_B$), as the logarithm of the ratio of two volumes in the $6N$-dimensional phase space. We are dealing with the accessible region of the phase space given the external constraints. However, $\Delta q$ and $\Delta p_{\sigma}$ do not represent the same variation.
In fact, $(\Delta q)^3$ is an extensive quantity measuring the total accessible volume in real space, the same for all particles $i=1,\dots,N$. Instead, $\Delta p_{\sigma}$ is an intensive quantity, proportional to the square root of the ratio $\frac{U}{N}$, measuring the standard deviation of the amount of dispersion in the quantities of movement $p_{j}$ (or, equivalently, the average of the $p_j^2$). If we look at the quantity $(\Delta p_{\sigma})^{3N}$ geometrically in the $3N$-dimensional $p$-part of the phase space, it represents the volume of the hypercube of side $\Delta p_{\sigma}=\sqrt{\frac{2mU}{3N}}$. Numerically, $\Delta p_{\sigma}$ has a microscopic value, $o\left(\frac{1}{\sqrt{N}}\right)$, compared to the macroscopic $\sqrt{2mU}$, supposed $o(1)$.

In order to recover Eq. (\ref{entro}) from Eq. (\ref{thermoentro2}), we should choose an 'absolute' reference state. The most natural choice is the $T=0$ quantum-mechanical ground state, to which we can attribute $S_{\rm ref}(T=0)=0$ by the third principle of thermodynamics, in keeping with the results of Section II. We need therefore to express the denominator $(\Delta p_{\sigma 0}\Delta q_0)^{3N}$ in the ground-state of the quantum-mechanical particle-in-a-box problem. We shall deal with it in Act III. 

Before, we remark that if, from Eq. (\ref{thermoentro2}), we blindly made the associations $(\Delta p_{\sigma}\Delta q)^{3N} \rightarrow \Omega(U,V,N)$ and $(\Delta p_{\sigma 0}\Delta q_0)^{3N} \rightarrow h^{3N} N!$, and considered the latter expression as the reference state of the statistical entropy, we would obtain a wrong expression (though extensive and with the correct dependence on variables $U$ and $V$), less than Eq. (\ref{sackur1}) by $Nk_B\ln\left(\frac{\pi e}{2}\right)^{\frac{3}{2}}$, as shown in Eq. (\ref{wrentro}). This proves that, starting from the thermodynamical entropy Eq. (\ref{thermoentro1}), we  do not obtain statistical entropy, Eq. (\ref{sackur1}) for a free gas, without some new physical ideas. We shall be back on this point in Act IV.

\vspace{0.3cm}

{\it Act III - Quantum ground state: the proper course.}
The well-known solution of the Schrödinger equation for a particle of mass $m$ in a one-dimensional box of length $L_0$ has eigenstates $\psi_n(x)=\sqrt{\frac{2}{L_0}}\sin\left(\frac{n\pi x}{L_0}\right)$ and corresponding eigenenergies $E_n=\frac{\pi^2\hbar^2n^2}{2mL_0^2}$, with $n=1,2,\dots$. Though we cannot define a phase space for quantum mechanics as the position coordinate ${\hat q}$ and the corresponding conjugate momentum ${\hat p}$ do not commute, Eq. (\ref{thermoentro2}) comes to our rescue, teaching us that, in order to define entropy, we rather need a $\Delta p_{0\sigma}\Delta q_0$-phase space, where $\Delta p_{0\sigma}$ is the square root of the single-particle energy eigenvalue. It is therefore perfectly defined in quantum mechanics in a basis of energy eigenvectors. 

Call $\Delta^{\pm} p_{0\sigma} = \pm\sqrt{ \langle p^2\rangle_{\psi_1} - \langle p\rangle_{\psi_1}^2}=\pm\sqrt{2mE_1}=\pm\frac{\pi\hbar}{L_0}=\pm\frac{h}{2L_0}$ and $\Delta q_0=L_0$. We used the property $\langle p\rangle_{\psi_n}=0$, $\forall \psi_n$, as the eigenstates are real, implying that there is no net motion, as already the case for Eq. (\ref{psigma}). The phase-space volume associated to the ground state is therefore $|\Delta^{\pm} p_{0\sigma}| \Delta q_0= \frac{h}{2L_0} L_0 =\frac{h}{2}$. We consider that both directions of the momentum are allowed by the symmetry of the macroscopic constraints\cite{comment2} and get the total $p$-phase-space volume $\Delta p_{0\sigma}=\Delta^+ p_{0\sigma}-(\Delta^- p_{0\sigma})=\frac{h}{L_0}$, so that $\Delta p_{0\sigma} \Delta q_0=h$.

If we generalized this calculation to the $3N$ degrees of freedom, we might be tempted to write $(\Delta p_{0\sigma}\Delta q_{0})^{3N}= h^{3N}$. However, this equality cannot be correct, because on the left we have an extensive quantity ($\Delta p_{0\sigma}$ is intensive and $\Delta q_{0}$ is extensive), whereas on the right we have lost the extensivity because in quantum mechanics $\Delta p_{0\sigma}$ is no more intensive but, rather, inversely proportional to $L_0$. 
It is therefore the introduction of the ground-state quantification condition in the quantity of movement, $|\Delta^{\pm} p_{0\sigma}|=\frac{h}{2L_0}$, that is responsible for the loss of the extensivity at the denominator and the necessity to find a way to cure it.
Though we know already that the solution to restore extensivity is the multiplication by $N!$, we choose not to impose it {\it ad hoc}, but rather invoke a general statistical principle, to be applied to both the numerator and denominator of Eq. (\ref{thermoentro2}), in order to obtain Eq. (\ref{entro}). We move to Act IV.

\vspace{0.3cm}

{\it Act IV -  Sum of microscopic configurations: deus ex machina.} 
We adopt the general principle that, in order to evaluate the statistical quantities, like $(\Delta p_{\sigma}\Delta q)^{3N}$ or $(\Delta p_{0\sigma}\Delta q_{0})^{3N}$, we must sum over all possible {\it classical} (i.e., distinguishable) microscopic configurations that are compatible with the macroscopic thermodynamical constraints (here the fixed values of $(U,V,N)$). These constraints are determined by the Hamiltonian of the system and the symmetry of its solutions. As known since Boltzmann, in the thermodynamic limit this method selects the most probable microscopic configurations. The new feature here is that we should apply the rule to both numerator ('target state') and denominator ('reference state') in the logarithm of Eq. (\ref{thermoentro2}).

\begin{figure}[ht!]
	\centering
\includegraphics[width=0.48\textwidth]{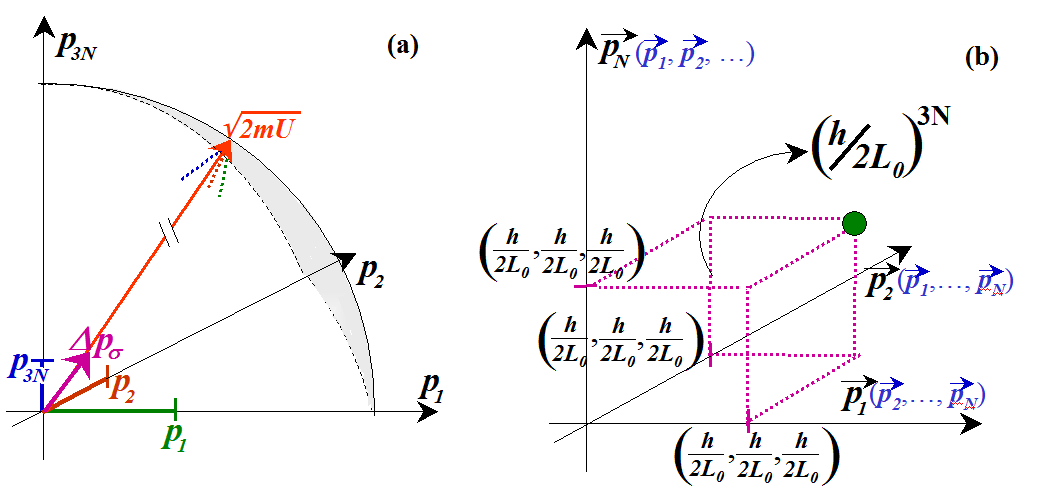}
	\caption{$3N$-dimensional $p$-phase space for (a) target state: the coordinates of the corner of the hypercube have 'microscopic' values ($o\left(\frac{1}{\sqrt{N}}\right)$), while the length of the diagonal has a 'macroscopic' value $o(1)$; an hyperspherical shell of highest volume-density is formed, when we allow all possible configurations compatible with the given constraints, from the corners of the hypercube of width $\Delta$. (b) reference state: as we limit to positive values, only one point in $\vec{p}$-space is represented. All $N!$ axes-permutations are allowed as the maximal subgroup of the $O(3N)$ symmetry of (a) compatible with the ground-state symmetry (with an abuse of notation, we have represented the axes by momentum vectors, see text).}
	\label{fig0}
\end{figure}

Start with the numerator: the allowed classical solutions keep the spherically symmetry of the Hamiltonian in the $p$-subspace and therefore the allowed microscopic configurations span the whole spherical shell. In this sense, the reason why Eq. (\ref{wrentro}) leads to a lower entropy than Eq. (\ref{sackur1}) is that it corresponds to a 'frozen' situation, where the hypercube of side $\Delta p_{\sigma}$ has not been allowed to pass through all microscopically accessible states. 
The spherical symmetry in the $3N$-dimensional $p$-phase space of the Hamiltonian implies that any rotoinversion of the hypercube around the origin is equally valid (which is the same as the rotoinversion of its diagonal, depicted in red in Fig. \ref{fig0}(a)). If we use the counter-intuitive property of high-dimensional hypervolumes that the most of the volume in the hypercube $(\Delta p_{\sigma})^{3N}$ is concentrated near the edge of the hypercube, it turns out that the most of the contribution to $(\Delta p_{\sigma})^{3N}$ comes from the spherical shell\cite{gibbsboltz} at the distance $\sqrt{2mU}$ from the origin.
Therefore, if we sum over all the possible rotations, we get : 

\begin{align}
\vspace{-0.2cm}
\hspace{0.5cm}\left(\Delta q \Delta p_{\sigma}\right)^{3N}\rightarrow V^N\int_{{\cal A}(p)} (dp)^{3N} = \Omega(U,V,N)
\label{stat0}
\end{align}

\noindent where ${\cal A}(p)$ is the phase-space region characterized by $U\leq \frac{1}{2m}\sum_{j=1}^{3N}p_j^2\leq U+\Delta$. An algebraic alternative to the present geometrical approach is reported in the Appendix.

The association of Eq. (\ref{stat0}) must be performed in parallel in the denominator $(\Delta q_{0}\Delta p_{0\sigma})^{3N}$.
So, we should sum over all microscopic configurations of the phase-space volume that are compatible with the macroscopic constraints $(U,V,N)$ and the symmetry of the solution of the Hamiltonian\cite{classref}. 
We know from Act III that the 'absolute' reference state corresponding to $T=0$ is the quantum-mechanical ground state of the particle-in-a-box Schrödinger equation. Because of its discreteness, this state is characterized by a different symmetry-group than the continuous $p$-space $O(3N)$ classical rotoinversion symmetry.
The only microscopic symmetry allowed for the quantum ground state (characterized by a point in the $3N$-dimensional $p$-subspace, as pictorially shown in Fig. \ref{fig0}(b), and by $2^N$ points in the full $p$-subspace) is the permutation symmetry of the $N$ momenta, leading to the same representative points. It is important to underline that the we are actually imposing the same {\it statistical} condition of maximal number of microstates compatible with the macroscopic constraints on both numerator and denominator. In fact, the permutation group $S_N$ of the axes of each ground-state momentum vector $\vec{p}_{{\rm GS}i}=(\frac{h}{2L_{0x}},\frac{h}{2L_{0y}},\frac{h}{2L_{0z}})$, $i=1,\dots,N$, is a subgroup of the rotoinversion group $O(3N)$, the maximal subgroup compatible with the discrete symmetry of the ground state. Therefore, the multiplication of $h^{3N}$ by $N!$ comes from the same request of {\it statistical} maximization, once the proper {\it quantum} constraints of the ground-state are taken into account.
We remark that the three directions in space are physically different, even if they may be numerically equal: $L_{0x}=L_{0y}=L_{0z}=L_{0}$. The quantization conditions are therefore not necessarily equivalent in the three directions and this is why the identity of the $\vec{p}_{{\rm GS}i}$ takes place at the level of the vectors (so, $N!$ permutations) and not at the level of the components (with $(3N)!$ permutations), as in the target state.
We shall comment further on this point in the Conclusions.
This leads to $N!$ microscopic configurations, each with weight $h^{3N}$.  

So, for the reference state we have:

\begin{align}
\vspace{-0.1cm}
\hspace{1cm} (\Delta p_{0\sigma}\Delta q_{0})^{3N} \rightarrow h^{3N}N!
\label{den}
\end{align}

\noindent and, by coupling Eqs. (\ref{stat0}) and (\ref{den}), we finally recover:

\begin{align}
\hspace{0.5cm} k_B \ln\left[ \left(\frac{\Delta p_{\sigma}\Delta q}{\Delta p_{0\sigma}\Delta q_0}\right)^{3N} \right] \longrightarrow  k_B\ln\frac{\Omega(U,V,N)}{h^{3N}N!}  
\label{thermoentro4}
\end{align}

\noindent which is the statistical entropy, Eq. (\ref{entro}). 

We conclude Act IV by underlying that, in the present approach the $N!$ in Eq. (\ref{entro}) is not justified by counting classical particles as indistinguishable. Rather, after applying the general principle of 'summing over all possible states' (equiprobable by symmetry) and exploiting the intrinsic differences of the high-temperature (continuous rotation symmetry) and zero-temperature (discrete permutation symmetry) phase spaces, we may say that we have counted the quantum labels of the reference state as distinguishable. 

Yet, this explains the case of Ar (identical masses), but not the case of Kr (different masses). We need to follow one more Act to close the story.

\vspace{0.3cm}

{\it Act V - Quantum vs. Statistical Indistinguishability: Catharsis.} 
Consider the classical Hamiltonian of an ideal gas of $N$ particles with different masses: $H=U=\sum_{i=1}^N\sum_{\lambda=1}^3\frac{p_{i\lambda}^2}{2m_i}$. As particles in the Hamiltonian are not identical, {\it quantum} indistinguishability is excluded by definition. Now, consider the average mass $m=\frac{1}{N}\sum_{i=1}^N m_i$ and the ratio $\alpha_i=\frac{m_i}{m}$. We have $2mU=\sum_{i=1}^N\sum_{\lambda=1}^3\frac{p_{i\lambda}^2}{\alpha_i}$. Perform the canonical transformation ${\tilde{p}}_{i\lambda}=\frac{p_{i\lambda}}{\sqrt{\alpha_i}}$ and ${\tilde{q}}_{i\lambda}=q_{i\lambda}\sqrt{\alpha_i}$, whose aim is to rescale the axes, so that the iso-energetic surface in the ${\tilde{p}}$-subspace is a sphere. The form of the Hamiltonian in the rescaled ${\tilde{p}}$-space is the same as in act II:  $H=U=\frac{1}{2m}\sum_{i=1}^N\sum_{\lambda=1}^3{\tilde{p}}_{i\lambda}^2$. We can therefore apply the same logical steps as in Act II and define $(\Delta {\tilde{p}}_{\sigma})^2=\frac{2mU}{3N}$, to end up with the ratio of the logarithm that becomes:  $\left(\frac{U^{\frac{3}{2}}V}{U_0^{\frac{3}{2}}V_0}\right)^N = \left(\frac{\Delta {\tilde{p}}_{\sigma}\Delta q}{\Delta {\tilde{p}}_{0\sigma}\Delta q_0}\right)^{3N}$.

The product $\Delta {\tilde{p}}_{0\sigma}\Delta q_0$ in the denominator does not lead to $h$, because the two variables are not conjugated. Rather, $(\Delta {\tilde{p}}_{0\sigma})^2=\left(\frac{h}{2L_0}\right)^2\left(\frac{1}{N}\sum_{j=1}^N\frac{1}{\alpha_j}\right)$, so that, including the factor $2^N$ for the denominator and the numerator, as in Acts III and IV, we find $\left(\Delta {\tilde{p}}_{0\sigma}\Delta {q}_{0}\right)^{3N} \rightarrow h^{3N} N!\left(\frac{1}{N}\sum_{j=1}^N\frac{1}{\alpha_j}\right)^{\frac{3N}{2}}$, where the $N!$ comes in for the same reasons as in Act IV (we remind that the permutation rule on ground-state momentum vectors is independent of the masses, whether identical or not - it only depends on the boundary conditions in the reference state).

We remark in passing that the canonical transformation employed here forces different boundary conditions in the ${\tilde{q}}$-space of the quantum-mechanical problem, as each single-particle eigenstate depends on the mass $m_i$ through the $\alpha_i$: ${\tilde{\psi}}_n^{(i)}(x)=\sqrt{\frac{2}{L_0\sqrt{\alpha_i}}}\sin\left(\frac{n\pi x}{L_0\sqrt{\alpha_i}}\right)$. For this reason, even if the particles in the transformed Hamiltonian appear as identical, the single-particle eigenstates are still {\it quantum-mechanically} distinguishable\cite{noteZ}. 

For the numerator, using the same rule as in Eq. (\ref{stat0}), we find: $\left(\Delta {\tilde{p}}_{\sigma}\Delta q\right)^{3N} \rightarrow V^N\int_{{\cal A}({\tilde{p}})} (d{\tilde{p}})^{3N} = \Omega(U,V,N) $, as ${\tilde{p}}$ is a dummy variable and that the volume of the phase space is defined in terms of the average mass, so that the latter expression is exactly the same as in Eq. (\ref{stat0}). The only difference with the gas of identical particles of mass $m$ seen in Act IV comes from the factor $\left(\frac{1}{N}\sum_{j=1}^N\frac{1}{\alpha_j}\right)^{\!\!\frac{3N}{2}}$ at the denominator.
If we apply this result to the case of Kr, we get a difference with respect to the Sackur-Tetrode expression of entropy with the average mass $m$ by $-\frac{3}{2}R\ln\left(\frac{1}{N}\sum_{j=1}^N\frac{1}{\alpha_j}\right)$. With the known isotopic masses ($m_1\simeq 83.91u$, $m_2\simeq 85.91u$, $m_3\simeq 81.91u$, $m_4\simeq 82.91u$ and $m_5\simeq 79.92u$), we get a correction of $- 0.003$ Jmol$^{-1}$K$^{-1}$, much below the experimental uncertainty.
This result provides the basis to include the division by $N!$ in both Kr and Ar calculations.

The latter calculation, however, does not explain, e.g., the case of isotopic separation\cite{schrod}. The point here is that, in order to separate the isotopes, the Hamiltonian cannot be the one proposed at the beginning of Act V: there must be an interaction connecting the different masses to the space variables (like the action of the gravitational potential in the method suggested by Lord Kelvin\cite{kelvin,darrigol}), otherwise the separation cannot be experimentally possible. Of course, the gravitational potential is also present in the experiment of Section II (where we divided by $N!$), and this leads to the following important statement: whether (non-identical) particles should be treated as distinguishable or indistinguishable depends on whether the experiment is conceived in such a way as to actually exploit the particle separation or not. We call this property {\it statistical} distinguishability/indistinguishability, respectively. In the case of Section II the coupling with the external (gravitational) field, though present, was not exploited in the experiment, so that all particles were subjected to the same boundary conditions. Therefore, all particles in the reference state were 'identical' with respect to the quantization of the $\vec{p}_{{\rm GS}i}$ and the entropy, using the maximization principle introduced in Act IV, is evaluated by dividing by $N!$: the particles are {\it statistically} indistinguishable. If, instead, the experiment is conceived in such a way as to exploit the particle separation through the external field\cite{schrod,kelvin,darrigol}, then the boundary conditions are different for all those particles characterized by the parameter allowing for the differentiation. For example, in the experiment suggested in Refs. [\onlinecite{schrod,kelvin}], the lightest particles are extracted by a hole, thereby changing the quantization conditions on $\vec{p}_{{\rm GS}i}$, in the reference state, for these particles (they are no more constrained by $L_0$). Therefore, their ground-state momenta are no more identical to those of the heavier particles that have not passed through the hole. These, say $N_1$, particles become {\it statistically} distinguishable and the denominator is separately factorized with respect to the $N_1!$ permutations of the $h^{3N_1}$ terms associated to the lighter particles and the $(N-N_1)!$ permutations of the $h^{3(N-N_1)}$ terms associated to the heavier particles: their statistical independence leads to $h^{3N}N_1!(N-N_1)!$. This is what we should measure for Kr if we were capable of separating, e.g., the lighter isotope from all the others (of course, if we were able to separate the 5 isotopes, we should divide by $\Pi_{i=1}^5(\gamma_i N)!$).

To conclude, $N$ particles are {\it statistically} indistinguishable if they have the same boundary conditions, leading to the same $N$ quantum-mechanical, ground-state, momentum vectors $\vec{p}_{{\rm GS}i}$ for the reference state, implying the division by $N!$ in Eq. (\ref{entro}) as a consequence of the maximization principle of Act IV. Conversely, $N_1$ of these particles are {\it statistically} distinguishable from the other $N-N_1$ if the boundary conditions for them are different (implicitly implying that the experimental parameters allow the distinction).
Of course, if a system is composed of identical particles (in the quantum sense), these particles are also {\it statistically} indistinguishable as there is no experimental way to differentiate them: in fact, in order to impose different boundary conditions between two classes of particles we need a physical parameter differentiating the two classes, which is by definition not the case for identical particles. But the converse is not true, as the previous example of Kr gas shows, i.e., {\it statistical} indistinguishability includes cases of non-identical particles, like also the example of colloids reported by Swendsen\cite{sw1}.  We shall further analyze this difference in the Conclusions.

\section{Is 'mixing-entropy' causally determined by particle mixing?}

What we have learned in the previous section is that the statistical entropy expressed by Eq. (\ref{entro}) is actually the ratio of two phase-space volumes subjected to the macroscopic thermodynamical constraints: the numerator $\Omega(U,V,N)$ is the phase-state volume of the 'target' state that we wish to evaluate, depending on the variables ($U$, $V$, $N$). The denominator, $h^{3N}N!$ is the phase-space volume of the reference state, which has been so chosen that it satisfies the third principle of thermodynamics. 

We shall now use this picture to explain what is usually\cite{kelvin,darrigol} called 'entropy of mixing' and show that mixing of particles is not the cause of the entropy change, that is rather determined by the different 'entropy-zero' determined by the experimental change of the boundary conditions.

\begin{figure}[ht!]
	\centering
	\includegraphics[width=0.5\textwidth]{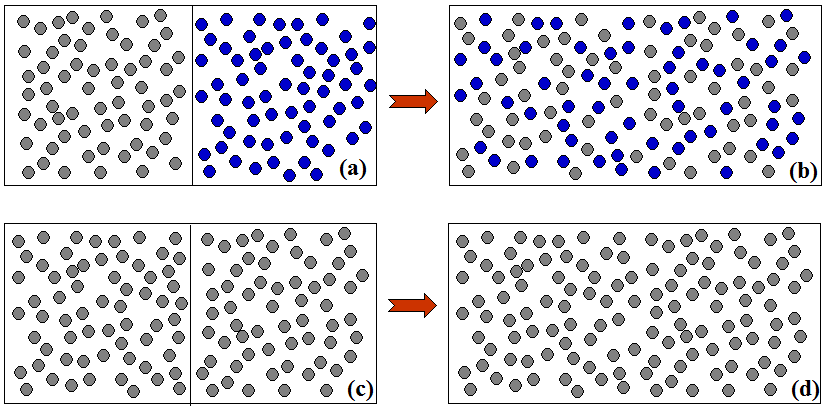}
	\caption{The so-called 'mixing-paradox': from (a) to (b) a partition is removed for distinguishable particles. From (c) to (d) a partition is removed for indistinguishable particles. In the first case we have an entropy difference of $Nk_B\ln 2$ and in the second case there is no entropy difference. As explained in the text, the increase from (a) to (b) is entirely determined by the increase in the available volume for each gas. The absence of entropy change from (c) to (d) is determined by the exactly canceling effects of the volume entropy-increase and the reference-state entropy-decrease.}
	\label{fig4}
\end{figure}

Consider, as an example, the 'standard' case depicted in Fig. \ref{fig4}. Panels (a) and (b) describe the thermodynamical transition where non-identical particles ($A$-kind and $B$-kind), initially separated by a partition in the two halves of the box, are then allowed to mix. Panels (c) and (d) describe the thermodynamical transition where identical particles (both of $A$-kind), initially separated by a partition in the two halves of the box, are then allowed to mix. In the former case, an entropy increase can be measured (or not, depending on whether we conceive the experiment so that it can be measured, or not), whereas the latter is, thermodynamically, a 'non-process' (the thermodynamical state does not change), with no entropy variation. In this section, we suppose that we can measure the entropy different for the non-identical particles in (b) panel.
In order to explain why the increase in entropy in this case is not due to the mixing of the particles, let's evaluate the entropy change in the two cases depicted in Fig. \ref{fig4}(a) and \ref{fig4}(b) with the interpretation of Eq. (\ref{entro}) developed in this work. For each figure, we evaluate the ratio of the phase volume of the 'target' (classical, high-temperature) state with respect to the phase-space volume of its corresponding 'reference' (quantum-mechanical, $T=0$) ground state.

Suppose for simplicity that, in both transitions, before we remove the partition, the two halves, of equal volume $V$, are in equilibrium at the pressure $p$, temperature $T$, implying that the number of particles in each half $N_A=N_B=\frac{N}{2}$, with $N$ the total number of particles in the total volume $2V$. 
Start with the entropy of the panel (a). It is the sum of the entropies in the left and in the right parts. The left part is: $S_L=k_B\ln\frac{\Omega_p^{(A)}V^{N_A}}{h^{3N_A}N_A!}$, where $\Omega_p$ is the classical phase-space volume for the $3N$-dimensional $p$-part, only. The right part is: $S_R=k_B\ln\frac{\Omega_p^{(B)}V^{N_B}}{h^{3N_B}N_B!}$. The entropy of panel (a) is $S_L+S_R=k_B\ln \frac{\Omega_p^{(A)}\Omega_p^{(B)}V^{N_A}V^{N_B}}{h^{3N_A+3N_B}N_A!N_B!}$.

Now evaluate entropy of panel (b). We have: $S_{LR}=k_B\ln \frac{\Omega_p^{(A)}\Omega_p^{(B)} (2V)^{N_A}(2V)^{N_B}}{h^{3(N_A+N_B)}N_A!N_B!}$, because the classical phase spaces $\Omega_p^{(A)}$ and $\Omega_p^{(B)}$ are independent for an ideal gas and therefore combine as a product. The real volume occupied by each gas $A$ and $B$ in panel (b) is doubled ($2V$), and, in the denominator, the permutations of $N_A$ and $N_B$ particles must be counted separately, because we have supposed the gases {\it statistically} distinguishable. 

How do we evaluate the entropy of the transformation from panel (a) to panel (b)? We should consider that both reference states, for panel (a) and for panel (b), have zero entropy by construction, as they represent their respective ground states. We can therefore imagine a full transformation from the reference state of (a), at $T=0$, whose entropy is zero, to the target state of (a), the one characterized by the entropy $S_L+S_R$, then towards the target state of (b), characterized by the entropy $S_{LR}$ and then to the reference state of (b), at $T=0$, whose entropy is again zero. So we have, for the whole transformation, $S_L+S_R + \Delta S_d - S_{LR}=0$, where $\Delta S_d$ is the entropy change from the target state of panel (a) to the target state of panel (b) that we wish to evaluate.
Therefore: 

\begin{align}
\Delta S_d &= S_{LR} - (S_L+S_R) = \nonumber \\
&\!\!= k_B\!\ln\!\left[\frac{\Omega_p^{(A)}\Omega_p^{(B)} (2V)^{N_A}(2V)^{N_B}}{h^{3(N_A+N_B)}N_A!N_B!} \frac{h^{3N_A}N_A!}{\Omega_p^{(A)}V^{N_A}}\frac{h^{3N_B}N_B!}{\Omega_p^{(B)}V^{N_B}}\right] \nonumber \\
& \!\!= k_B\ln 2^{N_A+N_B} = 2Nk_B \ln 2
\label{mix1}
\end{align}

We remark that the entropy increase is due to the {\it doubling of the volume} available to each species of particles and not their mixing.

Now turn to the transformation from panels (c) to panel (d). For panel (c) we have, analogously to the previous case: $S_L=k_B\ln\frac{\Omega_p^{(A)}V^{N_A}}{h^{3N_A}N_A!}$ and $S_{R2}=k_B\ln\frac{\Omega_p^{(A)}V^{N_A}}{h^{3N_A}N_A!}$. So, the entropy of panel (c) is $S_L+S_{R2}=k_B \ln \frac{(\Omega_p^{(A)})^2V^{2N_A}}{h^{6N_A}(N_A!)^2}$.
For panel (d), we have an important difference in the denominator compared to panel (b): the permutations of all $N_A$ particles must be counted together, because they are all {\it statistically} indistinguishable. This implies that the ground-state to which we attribute zero entropy for (d) panel of {\it statistically} indistinguishable particles is different than for (b) panel of {\it statistically} distinguishable particles. The result is: $S_{LR2}=k_B\ln \frac{(\Omega_p^{(A)})^2 (2V)^{2N_A}}{h^{3(2N_A)}(2N_A)!}$.
For the whole transformation going from the reference state of the (c) panel at $T=0$, to the target (c) state at finite $T$, then to the target (d) state at finite $T$ and finally to the reference (d) state at $T=0$, we have $S_L+S_{R2} + \Delta S_i - S_{LR2}=0$, where $\Delta S_i$ is the entropy change from the target state of panel (c) to the target state of panel (d), for {\it statistically} indistinguishable particles. Therefore, its value is (with $N=2N_A$):

\hspace{-0.5cm}
\begin{align}
\Delta S_i & = S_{LR2} - (S_L+S_{R2}) = \nonumber \\
&= k_B\ln\left[\frac{(\Omega_p^{(A)})^2  (2V)^{2N_A}}{h^{3(2N_A)}(2N_A)!} \frac{h^{3N_A}N_A!}{\Omega_p^{(A)}V^{N_A}}\frac{h^{3N_A}N_A!}{\Omega_p^{(A)}V^{N_A}}\right] \nonumber \\
& = k_B\ln \frac{2^{2N_A}N_A!N_A!}{(2N_A)!}  \nonumber \\
&= Nk_B\ln 2 - (2k_B\ln N_A! -k_B\ln(2N_A)!) \nonumber \\
& = Nk_B\ln 2 -Nk_B\ln 2 =0
\label{mix2}
\end{align}

\noindent where we used Stirling approximation $\ln n! \simeq n(\ln n-1)$ for the last equality of the second line. What we learn from Eq. (\ref{mix2}) is that in the case of mixing of indistinguishable particles there are two entropy contributions: one, positive, coming from the doubling of the volume, identical to the previous case of statistical distinguishable particles, and another, negative, coming from the statistical indistinguishability of the particles, which exactly balances the previous one, leading to zero entropy change.\cite{noteben}

In conclusion, the entropy increase in the case of  {\it statistically} distinguishable particles is not caused by their mixing, but by the volume increase for each particle species, as if the other species were not present.
Instead, when the particles are {\it statistically} indistinguishable, an extra entropy-reduction comes from the denominator so as to compensate for its increase due to the volume.

\section{Conclusions}

The main aim of the present paper is to gain a better insight into the origin of the $h^{3N}$ and of the $N!$ terms through the heuristic derivation of Eq. (\ref{entro}) from the Clausius' entropy presented in Section III. We used the property that the $3N$-dimensional, $p$-phase-space symmetries of the high-temperature, 'target', state and of the low-temperature, 'reference', state are different. The solution of the target state has the continuous\cite{quant} rotoinversion symmetry $O(3N)$, whereas the reference ground-state shows its quantum, discrete, nature, with 'only' $2^N$ points in the full $3N$-dimensional $p$-phase space, with coordinates $(\pm\frac{h}{2L_0},\dots,\pm\frac{h}{2L_0})$: in order to respect these discrete constraints, the rotoinversion symmetry $O(3N)$ invoked for the target state breaks into its maximal allowed subgroup $S_N$, the permutation group in $N$ dimensions, having $N!$ elements. 
We remark that, though in this article we stuck to the ideal-gas case for pedagogical purposes (calculations are simpler) the quantum-mechanical ground-state property of minimum momentum given by $(\frac{h}{2L_0},\dots,\frac{h}{2L_0})$ survives also in the case of interactions. Actually, it is a general geometric property of the quantum-mechanical ground state of a system of length $L_0$, whose wave-function goes to zero at the borders. In fact, we can continue the wave function by periodicity and expand it in Fourier series. The maximal wave-length is $2L_0$, thereby implying a minimal momentum $\frac{h}{2L_0}$. This is independent of whether masses are identical or not, as it is not a {\it dynamical} property, determined by the Hamiltonian, but a {\it geometric} property of the quantum ground-state, determined by the boundary conditions on the wave-function\cite{dyncond}. 
For this reason, even if we relax the ideal-gas condition, the property that the reference state is characterized by $N$ identical quantities of movement $\vec{p}_{{\rm GS}i}=(\frac{h}{2L_{0x}},\frac{h}{2L_{0y}},\frac{h}{2L_{0z}})$, $i=1,\dots,N$, is still valid. This gives a general validity to the expression of the denominator $h^{3N}N!$, which is based on this condition. In the last expression we have generalized the momentum quantization to the case $L_{0x}\neq L_{0y} \neq L_{0z}$, to explicitly show why the permutation group is $S_N$ and not $S_{3N}$: even if the three lengths are numerically equal, they still represent three different (orthogonal) directions in real space.  

Therefore, in the present interpretation, the origin of the $N!$ is both related to the {\it statistical} principle of maximization of the microscopically allowed reference states and to the {\it quantum} identity of the $N$ momenta in the ground state. Indeed, some relations between $N!$ and quantum mechanics had to be expected if we accept that the whole denominator $h^{3N}N!$ refers to the same state, as we needed its presence to correctly measure the Planck constant in Section II.
Yet we underline again that the quantum mechanical property referred to here is {\it not} the quantum indistinguishability of the particles, but rather the geometrical property of identity of all $\vec{p}_{{\rm GS}i}$, that only depends on the size of the three orthogonal directions of the volume and not on any kind of mass identity: this allows explaining also experiments on isotopes and colloids. So, with this interpretation, the often cited, unpleasant discontinuity in Eq. (\ref{entro}) with the nature of the particles (identical or not) does not exist.

As a personal remark, I would like to underline a possible source of confusion in dealing with entropy calculations, as two different questions might be posed, implying two different answers. The most common question is: 'How should we evaluate the entropy of a given system that is measured through a given experiment?'. It is this question that we tried to answer in this paper. 
However, another question might be: 'How should we evaluate the maximum amount of entropy that is obtainable for a given system by the most clever experiment we can imagine?'. The answer to this question is more tricky, because it appears to depend on the degree of knowledge that we have at the moment where we perform the measurement - but such a degree of knowledge can change in the future - and thereby the question falls in the category masterly described by Jaynes for the 'superalkali elements' Whifnium and Whafnium\cite{jaynes}. 
The best we can do to answer the second question, after adding the alert 'with our present degree of knowledge', is to take the example of the isotope separation described by Schrödinger\cite{schrod} but already proposed by Lord Kelvin\cite{kelvin} at a time where isotopes were not know. Suppose that Lord Kelvin had actually performed his experiment using Kr. Contrary to his expectation of zero entropy, valid in the case of same mass, he would have found an entropy production with possible extraction of a net work.
So, he might either have concluded to have found a violation of the entropy law expressed by Eq. (\ref{entro}), with $N!$, or if he - correctly - believed that the law could not be violated, he would have had an experimental proof of the existence of isotopes in Kr. This is why the second question, sometimes implicitly posed, is tricky, as taught to us by Jaynes\cite{jaynes}.
I conclude this point with a final remark that might be relevant for the interpretation of probabilities (frequentist, vs. bayesian): when we invoke the ergodic theorem, we usually admit a frequentist approach. Yet, the kind of microscopic states used in the reference state in this paper do not seem to be temporally connected and rather point towards the bayesian interpretation. 

To summarize, whereas the usual interpretation of Eq. (\ref{entro}) attributes the whole ratio $\frac{\Omega(U,V,N)}{h^{3N}N!}$ to the target state, I hope that the five-act Section III might have convinced you about the role of the quantity $h^{3N}N!$ in the denominator of the logarithm of Eq. (\ref{entro}): it represents the reference statistical entropy, in analogy with the quantity $(U_0^\frac{3}{2}V_0)^{N}$ in the thermodynamical entropy of Eqs. (\ref{thermoentro1}) and (\ref{thermoentro2}). Therefore, counting, e.g., in the case of Kr, with $N!$ or with $\Pi_{i=1}^5(\gamma_iN_i)!$ 'only' amounts to redefine the entropy zero, that depends on the specific experiment (boundary conditions) that we must describe.
Finally, the origin of the $N!$ is partly statistical, as a consequence of the statistical principle employed in Act IV of summing over all accessible microscopic states, and partly quantum: the quantum condition of identical ground-state momenta $\vec{p}_{{\rm GS}i}=(\frac{h}{2L_{0x}},\frac{h}{2L_{0y}},\frac{h}{2L_{0z}})$, $\forall i=1,\dots,N$, plays a fundamental role, as permutations would not have been a symmetry anymore if all ground-state momenta were different for all particles.
The latter property is only a consequence of the geometrical limitations, imposing a null wave function at the boundaries, and of our interpretation of the denominator as belonging, as a whole, to the $T=0$ reference state. Therefore, it is independent of the {\it quantum} indistinguishability and of particles identity.

\section*{Acknowledgement}

I thank C.R. Natoli for highlighting a methodological error in a previous version of the manuscript. I also thank the colleagues O. Emile and J. Emile, who read the previous version of the manuscript, pointing to some mistakes, and F. Greco, A. Simoni and J. Crassous for interesting discussions. Finally, this work would have never been written without the stimulating questions of my students at the ENS Rennes.

\vspace{0.5cm}

\appendix

\section{Extra calculations}

{\it Hypercube volume:} By comparing Eqs (\ref{thermoentro1}) and (\ref{thermoentro2}), we get for the numerator of the latter that: $(\Delta p_{\sigma} \Delta q)^{3N}=\left(\frac{2mU}{3N}\right)^{\frac{3N}{2}}V^N$, with the extra factor $2^{3N}$ due to the inclusion of the negative coordinates, as in Act IV. With the equivalent position $(\Delta p_{0\sigma} \Delta q_0)^{3N} \rightarrow \left(\frac{h}{2}\right)^{3N}N!$, as in Act III, we would have found for the entropy (with $S_0(N)=0$): 

\vspace{-0.2cm}

\begin{align}
S_{\sigma}(U,V,N)&=k_B\ln \left[ \frac{(8mU)^{\frac{3N}{2}}V^N}{(3N)^{\frac{3N}{2}}h^{3N}N!} \right] \nonumber \\
& = k_B N \left\{ \ln \left[\left( \frac{8mU}{3h^2N} \right)^{\frac{3}{2}} \frac{V}{N}  \right] +1 \right\}
\label{wrentro}
\end{align}

\vspace{-0.1cm}

\noindent with the usual Stirling approximation for the factorial. The difference with Eq. (\ref{sackur1}) is $S_{ST}-S_{\sigma}=k_BN\{\ln\left(\frac{\pi}{2}\right)^{\frac{3}{2}}  +\frac{3}{2}\}\simeq 2.18k_BN$. In terms of the molar entropy, it would have led to a difference of $s=2.18R\simeq 18.1$ JK$^{-1}$mol$^{-1}$.
This is the reason why we should discard the expression (\ref{thermoentro2}) as a possible origin of Eq. (\ref{entro}). 

{\it Algebraic interpretation of the maximization in Act IV:} We can use the multinomial theorem and write the following identities:

\vspace{-0.5cm}

\begin{align}
(2mU)^{\!\frac{3N}{2}} \!\!= \left(\sum\limits_{j=1}^{3N}\Delta p_j^2\right)^{\!\!\!\!\frac{3N}{2}} \!\!=\!\!\!\!\!\!\!\! \sum\limits_{\substack{k_i =1 \\ k_1+\dots+k_{3N}=\frac{3N}{2}}}^{3N}\!\!\!\!\!\!\!\!\frac{\left(\frac{3N}{2}\right)!}{k_1! \dots k_{3N}!}\prod\limits_{t=1}^{3N}{\Delta p_t^{2k_t}} \nonumber
\end{align}

\vspace{-0.2cm}

The geometric maximization of the microscopic classical configurations in the numerator of Act IV, leading to the spherical shell depicted in Fig. \ref{fig0}(a), finds its algebraic counterpart in the maximization of the coefficient $\frac{\left(\frac{3N}{2}\right)!}{k_1! \dots k_{3N}!}$ with respect to all the $k_i$ with the constraint that $\sum_{i=1}^{3N}k_i=\frac{3N}{2}$, a procedure justified when $N$ is a very big number, originally proposed by Boltzmann in a slightly different form. 
Even if in this case we cannot use the Stirling approximation on the $k_i!$ that are not necessarily big numbers, the result of the constrained maximization is $k_i=c$, a constant, $\forall i$. The constraint therefore imposes\cite{noteki} $k_i=\frac{1}{2}$ and leads to:

\vspace{-0.2cm}

\begin{align}
\label{stat2}
(2mU)^{\frac{3N}{2}} = \frac{\left(\frac{3N}{2}\right)!}{((\frac{1}{2})!)^{3N}}\prod\limits_{t=1}^{3N}{\Delta p_t^{2\frac{1}{2}}}  
& = \frac{2^{3N}\left(\!\!\frac{3N}{2}\!\!\right)!}{\pi^{\frac{3N}{2}}} \prod\limits_{t=1}^{3N}{|\Delta p_t|} \nonumber
\end{align}

\vspace{-0.2cm}

\noindent where we have used $(\frac{1}{2})! = \frac{\sqrt{\pi}}{2}$.
When we associate $(\Delta p_{\sigma 0}\Delta q_0)^{3N} \rightarrow h^{3N} N!$ at the denominator, the above association for the numerator leads to $(2|\Delta p_t|)^{3N}= (\Delta p_{\sigma})^{3N}=\left(\frac{4\pi e mU}{3N}\right)^{\frac{3N}{2}}$ and therefore to the Sackur-Tetrode equation (\ref{sackur1}), equivalent to Eq. (\ref{entro}) for an ideal gas.


\begin{thebibliography}{99}
\bibitem{noteB}
At the end of the XIX century, of course, Boltzmann could not suspect the presence of the Planck constant in Eq. (\ref{entro}).
\bibitem{comment}
In this work, we do not analyze the alternative, information-theoretical approach introduced by E.T. Jaynes\cite{physrev}. We shall comment on it in the Conclusions.
\bibitem{paulicomm}
The possibility to obtain an extensive entropy through a proper choice of the function ${\tilde{S}}_0(N)$ appearing in the integration of Clausius' formula was also remarked by Pauli\cite{pauli} and Jaynes.\cite{jaynes,jaynescomm}
\bibitem{jaynescomm}
Jaynes' Ref. \onlinecite{jaynes} has been such a source of inspiration for my own comprehension of statistical and thermodynamical entropy that I cannot but strongly recommend its reading.
\bibitem{extens}
This hypothesis presuppose extensivity, which is satisfied in experiments: entropy variations double when the amount of matter is doubled.
\bibitem{virial}
Usually, the comparison is performed with a virial correction to the ideal gas result. In our case, this correction, though non negligible, is rather low\cite{exposito}: $P\frac{dB}{dT}=0.50$ Jmol$^{-1}$K$^{-1}$. So, we should have rather compared $s^{\rm th}(T)$ to $s^e(T) + P\frac{dB}{dT} = 145.3 \pm 0.2$ Jmol$^{-1}$K$^{-1}$, still with an excellent agreement. We did not use this correction, not to overload the main text, as our conclusions are not affected.
\bibitem{notexpo}
We remark that the slight difference with Ref. [\onlinecite{exposito}], not relevant for our conclusions, is due to the data taken from the more recent version of Ref. [\onlinecite{handbook}].
\bibitem{noteXX}
Though the validity of our approach is limited to the ideal gas, we think that it is pedagogically useful for undergraduate students.
\bibitem{remark1}
The result $U=\frac{3}{2}Nk_BT$ is often derived from the generalized equipartition theorem,\cite{GET} whose demonstration is based on the explicit expression of statistical entropy (though only the numerator of Eq. (\ref{entro}) is actually needed). We highlight that the generalized equipartition theorem is however not necessary in this case, as shown in Ref. [\onlinecite{ma2}].
\bibitem{temp}
We remind that Refs. [\onlinecite{pauli,jaynes}] used $T$, intensive, instead of $U$, extensive, as independent variable. They were therefore concerned only by the volume variation, from $V_0$ to $V$, and not by internal energy variations.
\bibitem{notex} 
We do not care whether to define the variance with $\frac{1}{N}$ or with $\frac{1}{N-1}$ because $N > 10^{20}$.
\bibitem{comment2} 
The total macroscopic energy $U=NE_1$ is recovered with both $\Delta^+ p_{0\sigma}$ and $\Delta^- p_{0\sigma}$, so that the phase-space volume has to include both contributions.
\bibitem{gibbsboltz}
In the light of the recent controversy of the $\theta$-form vs. the $\delta$-form for Eq. (\ref{entro}), our approach seems to point to the $\theta$-form. However, it is not the aim of the article to discuss this point.
\bibitem{classref}
If the reference state were in the 'classical' region, it would have been legitimate to proceed as in the numerator: $(\Delta q_{0}\Delta p_{0\sigma})^{3N}\rightarrow V^N\int (dp)^{3N} \delta(2mU_0-\sum_{j=1}^{3N}p_j^2)\Delta$. Of course, this would not have led to Eq. (\ref{entro}).
\bibitem{noteben}
As written in Ref. [\onlinecite{bennaim}], the exact compensation of the two is a consequence of the Stirling approximation. 
\bibitem{quant}
Of course, this continuity is 'just' a consequence of the extremely high density of quantum-mechanical points for very high quantum numbers.
\bibitem{dyncond}
Actually, any geometrical conditions is a consequence of a dynamical interaction, in this case between the molecules of the gas and those of the walls. A Hamiltonian description is in principle possible if we include the degrees of freedom of the walls. It is not necessary here.
\bibitem{noteki}
In the multinomial coefficient, $k_i\in {\cal N}$, $\forall i$. So, we have to suppose its extension to fractional values.
\end{thebibliography}
\end{document}